\documentclass{JHEP3}
\usepackage{amsmath}
\usepackage{amssymb}
\usepackage{slashed}
\usepackage{graphicx}
\newcommand{\be}{\begin{equation}}
\newcommand{\ee}{\end{equation}}
\newcommand{\beqa}{\begin{eqnarray}}
\newcommand{\eeqa}{\end{eqnarray}}
\newcommand{\LL}{\mathcal{L}}
\newcommand{\cL}{\mathcal{L}}

\newcommand{\const}{\mathrm{const}}
\newcommand\e{\mathrm{e}}
\renewcommand\d{\partial}

\newcommand{\bseq}{\begin{subequations}}
\newcommand{\eseq}{\end{subequations}}

\title{Higgs inflation: consistency and generalisations}
\author{F. Bezrukov,${}^{a,c}$
  A. Magnin,${}^b$ M. Shaposhnikov${}^b$ and S. Sibiryakov${}^{b,c}$\\
  ${}^a$Max-Planck-Institut f\"ur Kernphysik,\\
   ~PO Box 103980, 69029 Heidelberg, Germany\\
  ${}^b$\'Ecole Polytechnique F\'ed\'erale de Lausanne,\\
   ~FSB/ITP/LPPC, CH-1015, Lausanne, Switzerland\\
  ${}^c$Institute for Nuclear Research of the
    Russian Academy of Sciences,\\
   ~60th October Anniversary Prospect, 7a, 117312 Moscow, Russia\\
  E-Mail: \email{fedor@ms2.inr.ac.ru}, \email{Amaury.Magnin@epfl.ch},
  \email{Mikhail.Shaposhnikov@epfl.ch}, \email{Sergey.Sibiryakov@cern.ch}
}

\abstract{
  We analyse the self-consistency of inflation in the Standard Model,
  where the Higgs field has a large non-minimal coupling to gravity.
  We determine the domain of energies in which this model represents a
  valid effective field theory as a function of the background Higgs
  field. This domain is bounded above by the cutoff scale which is
  found to be higher than the relevant dynamical scales throughout the
  whole history of the Universe, including the inflationary epoch and
  reheating. We
  present a systematic scheme to take 
 into account quantum loop corrections
  to the inflationary calculations within the framework of effective
  field theory. 
We discuss the additional assumptions that
must be satisfied by the ultra-violet completion of the theory to
allow connection between the parameters of the inflationary effective theory
and those describing the low-energy physics relevant for
the collider experiments. 
A class of generalisations of
  inflationary theories with similar properties 
is constructed.
}

\date{\today}

\begin{document}

%%%%%%%%%%%%%%%%%%%%%%%%%%%%%%%%%%%%%%%%%%%%%%%%%%%%%%%%%%%%%%%%%%%%%%%%
\section{Introduction}

It was proposed recently \cite{Bezrukov:2007ep} that the inflationary
expansion of the early Universe can be incorporated within the
Standard Model (SM). The SM already contains a particle---the Higgs
boson---with appropriate quantum numbers to play the role of the
inflaton. The key point of the Higgs-inflation is the  non-minimal
coupling of the Higgs field to gravity. Namely, the SM-gravity 
action is taken as
\begin{equation}
  \label{eq:TheAction}
  S = \int d^4x\sqrt{-g} \left(-
    \frac{M^2_P}{2}R - {\xi}H^\dagger{H}R + \cL_{SM}
  \right)
  \;,
\end{equation}
where $R$ is the scalar curvature, $H$ is the Higgs doublet, $M_P$ is
the Planck mass, $\cL_{SM}$ represents the SM Lagrangian and $\xi$ is
a new coupling constant (see
\cite{Spokoiny:1984bd,Futamase:1987ua,Salopek:1988qh,Fakir:1990eg,
  Kaiser:1994vs,Komatsu:1999mt,Tsujikawa:2004my} for previous studies of
inflation with non-minimally coupled scalar fields).
  
As found in \cite{Bezrukov:2007ep}, for the appropriate choice of
$\xi$ of the order $10^4$, the resulting model leads to successful
inflation, provides the graceful exit from it, and  produces the
spectrum of primordial fluctuations in good agreement with the
observational data.\footnote{For the minimal coupling to gravity
corresponding to $\xi=0$ an unacceptably large amplitude of primordial
inhomogeneities is generated for a realistic quartic Higgs
self-interaction \cite{Linde:1983gd}.}  Thus one arrives at an
economical scenario, where inflation does not require introduction of
any new degrees of freedom, with all necessary ingredients being
present in the SM.  This scenario  was further explored in
\cite{Barvinsky:2008ia,Bezrukov:2008ut,GarciaBellido:2008ab,
DeSimone:2008ei,Bezrukov:2008ej,Bezrukov:2009db,Barvinsky:2009fy}.

However, the self consistency of Higgs inflation was questioned in
\cite{Burgess:2009ea,Barbon:2009ya,Burgess:2010zq}.  It was pointed
out there that the operator describing non-minimal coupling, when
written in terms of canonically normalised fields, has dimension 5 and
is suppressed by the scale
\be
  \label{Lambda0}
  \Lambda_0=\frac{M_P}{\xi}
  \;.
\ee
This was interpreted as the ultra-violet (UV) cutoff, above which the
SM has to be replaced by a more fundamental theory. If true, this
would make the Higgs inflation ``unnatural''. Indeed,  for large
$\xi$, the scale $\Lambda_0$ is considerably lower than the Planck
mass. At the same time,  the value of the Hubble expansion rate is
close to $\Lambda_0$ during inflation, making the contribution of
unknown effects beyond the SM substantial \cite{Burgess:2009ea}.
Moreover, $\Lambda_0$ is much smaller than the value of the Higgs
field during inflation.  According to the standard lore, one would
argue that the action of the theory must be supplemented by other
higher-order operators suppressed by $\Lambda_0$, including additional
terms in the Higgs potential of the form
\be
  \label{bad}
  \frac{(H^\dagger H)^n}{\Lambda_0^{2n-4}}
\ee 
with $n\geq 3$.  These terms would  spoil the slow-roll regime.  Based
on these observations it was concluded in \cite{Barbon:2009ya} that
the validity of the Higgs inflation is very sensitive to the UV
completion of the theory.

Our aim in the present paper is to re-assess the self-consistency of
the Higgs inflation in order to disentangle the UV-sensitive aspects
of the model from those which can be analysed by means of an effective
field theory (EFT) description.  To make the analysis clear we
concentrate on a toy model of a single non-minimally coupled scalar
$\phi$, representing the radial mode of the Higgs field,
$H^\dagger{H}=2\phi^2$.  This allows to get rid of the complications
related to the gauge fields; effect of additional SM fields will be
discussed at the end of the paper.

We start by revisiting the calculation of the cutoff scale $\Lambda$,
to determine the region of validity of the theory with large
non-minimal coupling. We find that the domain of energies $E<\Lambda$,
were the model can be considered as a valid EFT \emph{depends} on the
background value of the scalar field. Its upper boundary 
coincides with
(\ref{Lambda0}) at $\phi=0$ and becomes higher for large background
values of $\phi$. We show that the background dependent cutoff is
parametrically higher than the energy scales characterising the
dynamics of the system throughout the whole history of the Universe.
In particular, during inflation, it coincides with the Planck mass (in
the Einstein frame\footnote{The {\it Einstein frame} is the frame
  where the non-minimal coupling between the inflaton and curvature is
eliminated in favour of essentially non-linear inflaton
self-interaction. It is related to the {\it Jordan frame}, where the
action (\ref{eq:TheAction}) is originally formulated, by a conformal
transformation, see Sec.~\ref{sec:Einstein}.}), which is much higher than the Hubble rate at that
moment. Thus the necessary requirements for the validity of the
semiclassical treatment of the model are satisfied. 

We next turn to the analysis of quantum corrections in the model. Of
course, the theory (\ref{eq:TheAction}) is non-renormalisable. In the
Jordan frame, the non-renormalisability is due to the coupling to
gravity, while in the Einstein frame it manifests itself in
essentially non-linear interactions of the scalar field. To remove the
divergences, one has to add an infinite number of counter-terms and the
corresponding finite terms with arbitrary coefficients. We show that
the counter-terms can be chosen in such a way that they respect the
classical symmetries of the model. The most important for us is the
{\em asymptotic} symmetry of the action under the shifts of the inflaton field
in the Einstein frame, corresponding to the asymptotic scale
invariance in the Jordan frame. This symmetry  exists in the domain
$\phi\to\infty$ relevant
for inflation. 

It is a well-known property of 
an approximate shift symmetry that it allows to preserve the
flatness of the inflaton potential
under radiative corrections \cite{Freese:1990rb}. 
In this paper we develop 
 a systematic scheme
which takes into account breaking of the symmetry order
by order in perturbation theory.
It leads to the classification of the
operators generated by quantum corrections according to their
order in the parameter that controls breaking of the shift invariance
at finite values of the field. This leads to an EFT description,
close in spirit to that proposed in \cite{Cheung:2007st,Weinberg:2008hq},
where the expansion is effectively controlled by the inflationary
slow-roll parameters. Importantly, the asymptotic symmetry
precludes the generation of counter-terms of the type (\ref{bad}). 
So, the Higgs inflation is self-consistent and ``natural''.
Note that a similar conclusion was achieved recently in
\cite{Ferrara:2010in}.

It is worth stressing that in this paper we take a `minimalistic' 
attitude to the
self-consistency issue. Namely, we
consider only those quantum corrections which are forced by the 
inflationary theory itself. In particular, the asymptotic
shift symmetry which is the property of the inflationary dynamics is
{\em assumed} to be valid also at the level of the UV-complete
theory. Alternatively, this can be considered as a restriction on the
UV-completion which must be satisfied for our results to remain in
force. We do not address the origin of the asymptotic shift symmetry
in the UV theory.  
This question has been discussed recently in
\cite{Baumann:2010ys} (see also references therein).

The EFT approach to inflation that we develop in this paper is
general. Besides the model (\ref{eq:TheAction}), we show how it
can be applied to
a wide class of inflationary Lagrangians with asymptotic shift
symmetry. 

Finally, we discuss under which conditions the parameters of the
inflationary EFT can be connected to those describing the low-energy
physics relevant for collider experiments. We find that this
connection 
 is \emph{sensitive} to the
details of the UV completion.  Thus, no relation between these
parameters can be established in general, without specific assumptions
about the physics beyond the cutoff. We determine explicitly the
requirements to the UV completion, which lead to the relation between
the low-energy and inflationary domains. We discuss, in particular,
the sensitivity of the connection of the Higgs boson mass and the
inflationary parameters
\cite{Barvinsky:2008ia,Bezrukov:2008ej,DeSimone:2008ei,
Bezrukov:2009db,Barvinsky:2009fy,Barvinsky:2009ii} to UV physics.

The paper is organised as follows.  In Sec.~\ref{sec:scale} we perform
the calculation of the background dependent cutoff.  In
Sec.~\ref{sec:quantum} we address the structure of quantum corrections
and develop the effective field theory for the inflationary epoch.
Section~\ref{sec:general} discusses the generalisations of the
Higgs-inflation to a wide class of inflationary Lagrangians with
asymptotic shift symmetry.  Section~\ref{sec:conclusions} is devoted
to conclusions.  Appendices contain analysis of the model with
addition of fermions and gauge bosons.

%%%%%%%%%%%%%%%%%%%%%%%%%%%%%%%%%%%%%%%%%%%%%%%%%%%%%%%%%%%%%%%%%%%%%%%%
\section{The cutoff scale revisited}
\label{sec:scale}
Let us start by discussing the definition of the cutoff scale.  The
useful criterion for the validity of  perturbation theory is the tree
unitarity \cite{Cornwall:1974km}, which means that all $N$-particle
tree amplitudes $M_N$ have at most a high-energy behaviour
\be
  \label{reg}
  M_N \propto E^{4-N}
  \;,
\ee
where $E$ is the energy. This is the case in renormalisable theories
which thus can be considered as fundamental theories valid at arbitrary
momenta (we leave aside the issue of Landau poles which, if present,
occur at exponentially high energies and are irrelevant for our
discussion). If instead the tree amplitudes grow with energy or fall
slower than (\ref{reg}), the perturbation theory fails at some energy
$\Lambda$, which can be called an ultra-violet cutoff.  Whether the
theory gets inconsistent at energies higher than $\Lambda$, or just
enters into a strongly interacting phase, can not be deduced a priori.
In any event, the theory is only predictive with the use of
traditional perturbative methods at energies $E<\Lambda$.  Thus we
arrive at the following definition of the cutoff scale $\Lambda$.
Compute all tree amplitudes with $N$ particles and find the energy
$\Lambda_N$ at which the unitarity bound in each of them is violated.
Then define the cutoff as\footnote{Some care is needed in applying
this definition.  One should check that it does not put too much
weight into the multiparticle amplitudes with $N\gg 1$, for which
conventional perturbation theory breaks down even in the case of
renormalisable field theories, see e.g.\ \cite{Voloshin:1990mz}.}
$\Lambda=\min_N\Lambda_N$.

The cutoff scale is not just a number. It depends, in general, on
background bosonic field(s). For example, the cutoff of the 4-fermion
low energy Fermi theory of weak interactions is proportional to the
expectation value of the Higgs field. So, to define the region of
validity of the theory (\ref{eq:TheAction}), one should fix the 
background and consider the asymptotic high energy behaviour of tree
$N$ particle amplitudes. This is exactly what we have to do to
understand the viability of the Higgs inflation, because during inflationary
evolution of the Universe, the system is not described by its
perturbations about the vacuum solution, but rather by excitations
above some classical background. Thus the fields are naturally divided
in the slowly varying classical part and excitations on it
\begin{equation}
  \Phi(\mathbf{x},t)=\bar\Phi(t)+\delta\Phi(\mathbf{x},t)
  \;,
\end{equation}
where $\Phi$ stands for the generic set of fields in the theory
(inflationary scalars, gravitational metric, etc.). 
The perturbations relevant for the cutoff
determination have high frequencies corresponding to short time
scales. These are much shorter than the typical time scale of the
background evolution.
Thus the background 
can be approximated as static with a good
accuracy.

Instead of actually calculating the $N$ particle amplitudes, we will
estimate the cutoff by power counting of the operators, present in the
expansion of the action in $\delta\Phi$.  That is, if we have
operators with dimension larger than four, divided by some scale
$\Lambda_{(n)}$,
\begin{equation}
  \label{eq:OnLn}
  \frac{\mathcal{O}_{(n)}(\delta\Phi)}{[\Lambda_{(n)}(\bar\Phi)]^{n-4}}
  \;,
\end{equation}
then we expect the tree level unitarity to be violated in some
amplitudes at the scale $\Lambda_{(n)}$.  Clearly, the cutoff scale
determined in this way does depend  on the background values of the
fields, which is indicated explicitly in (\ref{eq:OnLn}).  Some
non-trivial cancellations may alter the result raising the unitarity
violation scale.  Thus, strictly speaking, this approach provides a
lower estimate of the cutoff.  We will neglect possible cancellations
in what follows: after all, having the lower bound on the cutoff is
enough for our purposes.  Besides, we are going to see that the simple
power counting estimates for the cutoff agree in two different
representations of the theory which favours the identification of
these estimates as the true value of the cutoff.

%%%%%%%%%%%%%%%%%%%%%%%%%%%%%%%%%%%%%%%%%%%%%%%%%%%%%%%%%%%%%%%%
\subsection{Cutoff in the Jordan frame}
\label{sec:Jordan}

We now turn to the calculation of the cutoff scale for the
Higgs-inflation model.  We start by performing the analysis in the
Jordan frame where the model was originally formulated.  Throughout
the main part of  the paper we work with a toy model of a single real
scalar field with non-minimal coupling to gravity (the effects of
matter fermions and gauge bosons  will be discussed later).  The
action of the model is
\be
  \label{Lagr1}
  S = \int d^4x\sqrt{-g}\left( -\frac{M^2+\xi\phi^2}{2}R
    + \frac{(\d_\mu\phi)^2}{2} - \frac{\lambda\phi^4}{4}
  \right)
  \;.
\ee
We have neglected the mass term for $\phi$ since it is not important
at large values of the field.  This is the scalar part of the action
for the SM Higgs inflation in the unitary gauge
\cite{Bezrukov:2007ep}.   The way inflation proceeds in this model is
described in detail in \cite{Bezrukov:2007ep}, while the reheating was
studied in \cite{Bezrukov:2008ut,GarciaBellido:2008ab}. 
For our  purposes we need the
following results. The non-minimal coupling with curvature modifies
the kinetic term for the scalar field for large values of the field,
leading to the slow-roll evolution even with relatively large quartic
coupling constant $\lambda$. Normalisation of the primordial density
fluctuations to the observed value fixes the relation between
$\lambda$ and non-minimal coupling $\xi$
\begin{equation}
  \label{eq:9}
  \xi \simeq 47000\sqrt{\lambda}
  \;.
\end{equation}
This means, that if $\lambda$ is not very small, as in the case of the
SM Higgs boson, $\xi$ should be rather large. In this limit 
inflation happens for the values of the Jordan frame field
$\phi>\phi_{\mathrm{END}}\simeq M/\sqrt{\xi}$.

To obtain the scale of tree-level unitarity violation we expand the
metric and the scalar around their background values,
\begin{align}
  \label{gexp}
  g_{\mu\nu} &= \bar g_{\mu\nu}+h_{\mu\nu}
  \;,\\
  \label{phiexp}
  \phi &= \bar\phi+\delta\phi
  \;.
\end{align}
The quadratic Lagrangian for the excitations has the form,
\be
  \label{Lagr2}
  \begin{split}
    \LL^{(2)} =
    -\frac{M_P^2+\xi\bar\phi^2}{8}\big(h^{\mu\nu}\Box h_{\mu\nu}
    +2\d_\nu h^{\mu\nu}\d^\rho h_{\mu\rho}-2\d_\nu h^{\mu\nu}\d_\mu h
    -h\Box h\big) \\
    +\frac{1}{2}(\d_\mu\delta\phi)^2
    +\xi\bar\phi\big(
      \Box h-\d_\lambda\d_\rho h^{\lambda\rho}
    \big)\,\delta\phi
    \;,
  \end{split}
\ee
where $h=h^\mu_\mu$.  We retained here only the terms with two
derivatives of the excitations as they determine the UV behaviour of
the scattering amplitudes and hence the unitarity violation scale.
Notice, that in the nontrivial background there is a large kinetic
mixing between the trace of the metric and the scalar perturbations
\cite{Barvinsky:2008ia,DeSimone:2008ei,Barvinsky:2009fy,Barvinsky:2009ii}. 
The change of variables
\begin{align}
  \label{chisubs}
  \delta\phi &=
  \sqrt{\frac{M^2_P+\xi\bar\phi^2}{M^2_P+\xi\bar\phi^2+6\xi^2\bar\phi^2}}
  \,\delta\hat\phi
  \;,\\
  \label{hsubs}
  h_{\mu\nu} &= \frac{1}{\sqrt{M^2_P+\xi\bar\phi^2}} \,\hat h_{\mu\nu}
  -\frac{2\xi\bar\phi}{\sqrt{(M^2_P+\xi\bar\phi^2)(M^2_P+\xi\bar\phi^2+
      6\xi^2\bar\phi^2)}} \, \bar{g}_{\mu\nu}\,\delta\hat\phi
\end{align}
diagonalises the kinetic term.  The unitarity violation scale is now
read out of the operators with dimension higher than four.  The
leading operator is the cubic scalar--graviton interaction
$\xi(\delta\phi)^2\Box h$, which in terms of the canonically
normalised variables has the form,
\be
  \label{dim3cutoffoprator}
  \frac{\xi\sqrt{M^2_P+\xi\bar\phi^2}}{
    M^2_P+\xi\bar\phi^2+
    6\xi^2\bar\phi^2}
  (\delta\hat\phi)^2 \Box\hat h
  \;.
\ee
The cutoff is identified as the inverse of the coefficient in this
operator,
\be
  \label{cutoff}
  \Lambda^J(\bar\phi) = \frac{M^2_P+\xi\bar\phi^2+
    6\xi^2\bar\phi^2}{\xi\sqrt{M^2_P+\xi\bar\phi^2}}
  \;,
\ee
where the superscript $J$ reminds that this cutoff is obtained in the
Jordan frame.  The expression for the cutoff simplifies in three
regions of background fields:

\begin{itemize}
\item $\bar\phi\ll M_P/\xi$, low field region.  This region
  corresponds to the present-day Universe.  The cutoff is
  \be
    \label{cutoffsmall}
    \Lambda^J \simeq \frac{M_P}{\xi}
    \;.
  \ee
  This coincides with the zero background result of
  \cite{Burgess:2009ea,Barbon:2009ya,Burgess:2010zq}.  It is smaller
  than the Planck mass, but for the observationally required value of
  $\xi\sim10^4$ it is still way above the reach of collider experiments.
\item $M_P/\xi\ll\bar\phi\ll M_P/\sqrt{\xi}$, the intermediate region
  (relevant for reheating, see
  \cite{Bezrukov:2008ut,GarciaBellido:2008ab}). The cutoff
  scales as
  \be
    \label{cutoffint}
    \Lambda^J \simeq \frac{\xi\bar\phi^2}{M_P}
    \;.
  \ee
  It is still below the Planck mass but
  starts to grow.
\item $\bar\phi\gg M_P/\sqrt{\xi}$, large fields (inflationary period).  The
  cutoff becomes
  \be
    \label{cutofflarge}
    \Lambda^J \simeq \sqrt{\xi}\bar\phi
    \;.
  \ee
  Note that this coincides with the cutoff in the gravitational
  sector. The latter is
  given by the effective Planck mass defined as the coefficient in
  front of the $R$ term in the Lagrangian,
  $M_P^{\mathrm{eff}}=\sqrt{M^2_P+\xi\bar\phi^2}$.
\end{itemize}
The behaviour of the cutoff is illustrated in Fig.~\ref{fig:1}.

\FIGURE[ht]{
\label{fig:1}
\centerline{\includegraphics[width=0.6\textwidth]{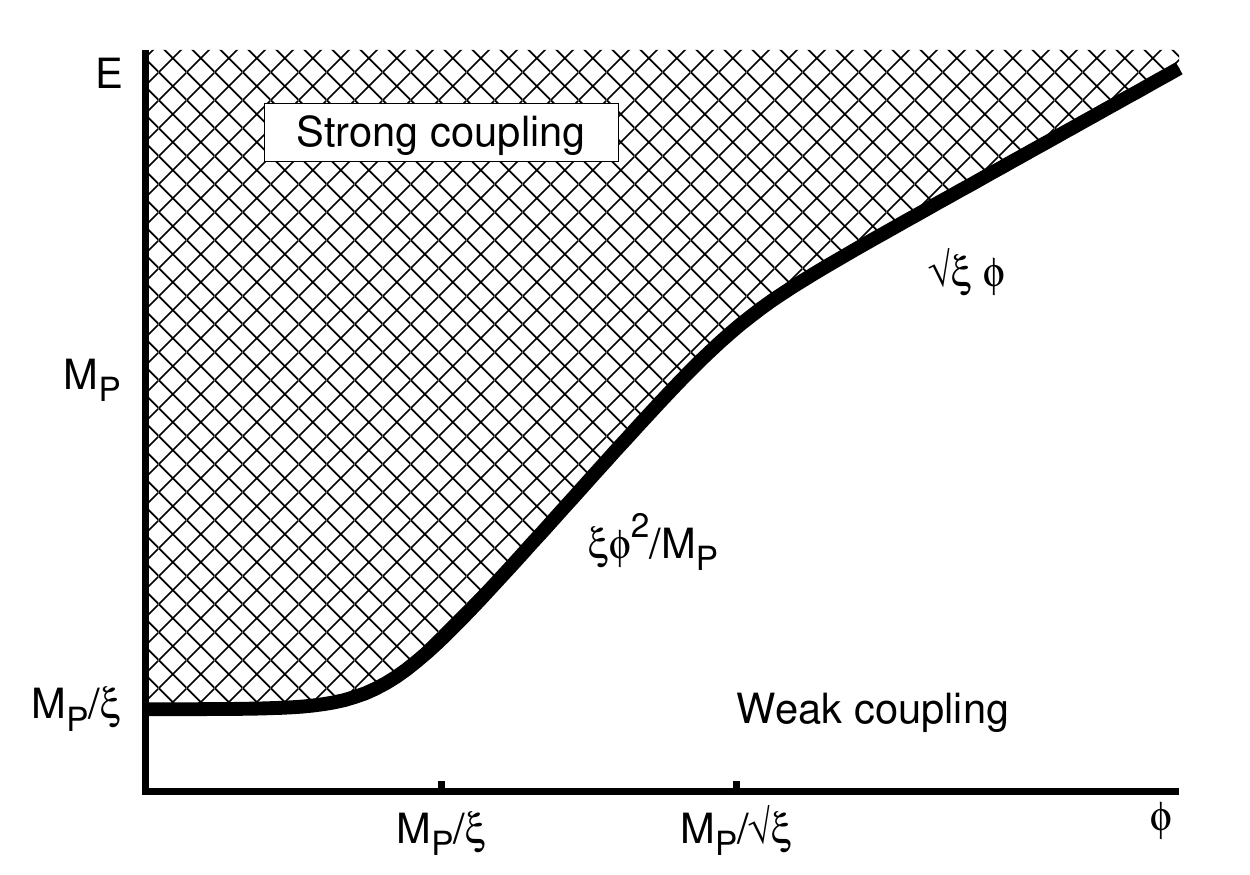}}
\caption{The dependence of the Jordan frame cutoff on 
the background value of the
  inflaton field in log-log scale. The effective field theory
  description is applicable at energies below the thick curve.}
}

%%%%%%%%%%%%%%%%%%%%%%%%%%%%%%%%%%%%%%%%%%%%%%%%%%%%%%%%%%%%%%%%
\subsection{Cutoff in the Einstein frame}
\label{sec:Einstein}

It is instructive to repeat the calculation of the cutoff scale in the
Einstein frame.  In this frame the gravitational part of the action
coincides with that of the usual Einstein's gravity, while all
non-trivial interactions are moved exclusively into the scalar sector.
This makes the analysis in the Einstein frame conceptually simpler and
we will work in this frame from now on.  The Jordan and Einstein
frames are related by the conformal transformation
\begin{equation}
  \label{conftrans}
  g_{\mu\nu}=\Omega^{-2}\tilde g_{\mu\nu}
  \;,
\end{equation}
with the conformal factor 
\be
  \label{Omega}
  \Omega^2=1+\frac{\xi\phi^2}{M^2_P}
  \;.
\ee
Substituting this into (\ref{Lagr1}) we obtain,
\be
  \label{Einstlagr}
  S=\int d^4x\sqrt{-\tilde g}\left( -\frac{M^2_P}{2}\tilde R
    + \frac{M_P^2(M_P^2+(6\xi+1)\xi\phi^2)}{(M_P^2+\xi\phi^2)^2}
      \frac{(\d_\mu\phi)^2}{2}
    - \frac{\lambda M_P^4\phi^4}{4(M_P^2+\xi\phi^2)^2} \right)
  \;,
\ee
where tilde denotes the geometrical quantities calculated in the
Einstein frame.  Note that the scalar potential flattens out and tends
to a constant at large $\phi\gg M_P/\sqrt{\xi}$.  This is the origin
of the slow-roll inflation in the Einstein frame picture
\cite{Bezrukov:2007ep}.

To proceed we have two options.  One is to work directly with the
field $\phi$.  Then, expanding it around the background, one reads out
the cutoff as the scale suppressing higher-order interactions
appearing from the kinetic term.  We take
another route and perform the field redefinition which casts the
kinetic term into the canonical form.  This is achieved by introducing
field $\chi$ related to $\phi$ by
\be
  \label{varphiphi}
  \frac{d\chi}{d\phi} =
  \frac{M_P\sqrt{M_P^2+(6\xi+1)\xi\phi^2}}{M_P^2+\xi\phi^2}
  \;.
\ee
In terms of this new field all non-linearities are moved into the
potential
\begin{equation}
  \label{eq:Efpotential}
  U(\chi)=\frac{\lambda M_P^4\phi(\chi)^4}{4(M_P^2+\xi\phi(\chi)^2)^2}
  \;.
\end{equation}
Expanding as usual above the background,
\begin{equation}
  \label{split}
  \chi=\bar\chi+\delta\chi
  \;,
\end{equation}
one calculates the Taylor expansion of the potential
\begin{equation}
  \label{eq:6}
  U(\bar\chi+\delta\chi) = U(\bar\chi) +
    \sum_{n=1}^\infty\frac{1}{n!}\frac{d^nU}{d\chi^n}(\delta\chi)^n
  \;.
\end{equation}
In this expansion the operators with $n>4$ will contribute to the
$n$-particle scattering amplitudes in a non-unitary way starting from
the energy scale
\begin{equation}
  \label{eq:7}
  \Lambda_{(n)} \sim \left(\frac{d^nU}{d\chi^n}\right)^{-\frac{1}{n-4}}
  \;.
\end{equation}
Let us analyse this expansion in the same three regions of the
background field as before.

\begin{itemize}
\item $\bar\phi,\bar\chi\ll M_P/\xi$.  In this case inverting
  (\ref{varphiphi}) one finds
  \be
    \label{eq:philowchiexpansion}
    \phi(\chi)=\chi\left(1+\sum_{k=1}^\infty c_k
      \left(\frac{\xi^2\chi^2}{M_P^2}\right)^k\right)
    \;,
  \ee
  where $c_k$ are numerical coefficients.  Substituting this into
  (\ref{eq:Efpotential}), (\ref{eq:7}) one obtains,
  \begin{equation}
    \label{eq:8}
    \Lambda_{(n)} \sim \frac{M_P}{\xi} \lambda^{-1/(n-4)}
    \;
  \end{equation}
  for even $n$.  This differs from the Jordan frame expression
  (\ref{cutoffsmall}) by a power of $\lambda$.  However, this factor
  tends to unity for operators of higher dimension.  If $\lambda$ is
  not too small, the factor becomes close to one already for
  moderately large $n$.  Thus we omit it in the determination of the
  cutoff and arrive at the expression (\ref{cutoffsmall}).
\item Intermediate region $M_P/\xi\ll\bar\phi\ll M_P/\sqrt{\xi}$. For the
Einstein frame field this corresponds to the region $M_P/\xi\ll\bar\chi\ll
M_P$.  The calculation of the cutoff using (\ref{eq:7}) requires
calculation of higher derivatives of the composite function
$U(\phi(\chi))$ using (\ref{eq:Efpotential}), (\ref{varphiphi}). 
Direct calculation is rather tedious\footnote{Note that in this
calculation one cannot use the approximate expression for the
potential (\ref{pot1}) as it does not lead to
the required precision for higher derivatives.} (though obviously
possible), so we will sketch here a simplified way to retain the leading
behaviour of the higher derivatives for this region of 
$\bar\phi$ values.  It
is convenient to switch from $\phi$ to the variable $z=\phi^2/M_P^2$
(i.e.\ $\xi z\ll1\ll\xi^2 z$).  Then we have
  \begin{equation}
    f(z)\equiv \frac{dz}{d\chi} \simeq
    \frac{2}{\sqrt{6}\xi M_P}\left(
      1+\xi z-\frac{1}{12\xi^2z}+\dotsb
    \right)
    \;,
  \end{equation}
  where the dots represent higher order corrections in $\xi z$ or
  $(\xi^2 z)^{-1}$. Using the chain rule one obtains the $n$-th derivative
  of the potential
  \begin{equation}
    \frac{d^n U}{d\chi^n} =
    \sum \const\cdot U^{(m)}(z) \prod_{j=0}^{n-1}\big(f^{(j)}(z)\big)^{m_j}
    \;.
  \end{equation}
Here $U^{(k)}(z)$, $f^{(k)}(z)$ denote $k$-th derivatives with respect
to $z$ and the sum runs over all sets of integers $m\geq 1$, $m_j\geq
0$ subject to the restrictions
  $m+m_1+2m_2+\dotsb+(n-1)m_{n-1}=n$,~ $m_0+\dotsb+m_{n-1}=n$.  The
  leading term in this sum for large $n$ is
  \begin{equation}
    U'(z)\big(f(z)\big)^{n-1}f^{(n-1)}(z) \propto
    \frac{\lambda}{\xi^{n+2}M_P^{n-4}(\phi^2/M_P^2)^{n-1}}
    \;.
  \end{equation}
  Using (\ref{eq:7}) we immediately 
get\footnote{To be precise, the next-to-leading
    term is $U''(z)\big(f(z)\big)^2\big(f'(z)\big)^{n-2}
\propto\lambda/(\xi^2M_P^{n-4})$.  It can be
    neglected in the region $M_P/\xi\ll\bar\phi\ll M_P/\sqrt{\xi}$ only
    for relatively large $n$.  For small $n$ the estimate
    (\ref{eq:cutoffintE}) is 
    true only for the lower part of the intermediate region.}
  \begin{equation}
    \label{eq:cutoffintE}
    \Lambda_{(n)} \sim \frac{\bar\phi^2\xi}{M_P}\cdot
\bigg[\frac{\xi^6\bar\phi^6}{\lambda M_P^6}\bigg]^{1/(n-4)}
    \;,
  \end{equation}
which reproduces 
(\ref{cutoffint}) at $n\to\infty$.

It is worth mentioning that a
   technically simpler way to obtain the cutoff in this region is to
  analyse the theory with additional fermions.  Being out of the
  main line of the article this analysis is given in Appendix
  \ref{app:fermions}.
\item $\bar\phi\gg M_P/\sqrt{\xi}$, inflationary region.  For the
  Einstein frame field this corresponds to $\bar\chi\gg M_P$.  In this
  region one can neglect $M_P^2$ in the numerator of
  (\ref{varphiphi}) which yields simple analytic solution
  \be
    \label{chiphi2}
    1+\frac{\xi\phi^2}{M^2_P}\simeq\exp\left(\frac{2\chi}{\sqrt{6}M_P}\right)
    \;.
  \ee  
  Substituting this into (\ref{eq:Efpotential}) we obtain the
  expression for the potential
  \be
    \label{pot1}
    U(\chi)=\frac{\lambda M^4_P}{4\xi^2}
    \left(1-\exp\left(-\frac{2\chi}{\sqrt{6}M_P}\right)\right)^{2}
    \;.
  \ee
  Expanding it around a background we get a series of interactions
  which have the form
  \be
    \frac{\lambda M^4_P}{\xi^2}\exp\left(-\frac{2\bar\chi}{\sqrt{6}M_P}\right)
    \;\frac{(\delta\chi)^n}{M^n_P}
    \;.
  \ee
  Note that these interactions contain an overall exponential
  suppression.  However, for any fixed $\bar\chi$ the energy cutoff
  scale goes to
  \begin{equation}
    \label{EinsLambdainfl}
    \Lambda\sim M_P
  \end{equation}
  for large $n$.  To compare this result with that in the Jordan frame
  (\ref{cutofflarge}) we should rescale it using the conformal factor
  (\ref{Omega}).  The latter is approximately equal to
  $\sqrt{\xi}\bar\phi/M_P$ for large field.  Multiplied by
  (\ref{EinsLambdainfl}) this reproduces (\ref{cutofflarge}).  Let us
  note again, that in the inflationary region the cutoff in the scalar
  sector coincides with the gravitational cutoff (which is just $M_P$
  in the Einstein frame or $\sqrt{\xi}\bar\phi$ in the Jordan frame).
\end{itemize}

%%%%%%%%%%%%%%%%%%%%%%%%%%%%%%%%%%%%%%%%%%%%%%%%%%%%%%%%%%%%%%%%
\subsection{Cutoff and energy scales in the early Universe}
\label{sec:Universescales}

Let us compare the background-dependent cutoff derived above with the
characteristic energy scales during the evolution of the Universe.  We
perform the discussion in the Einstein frame.  At the inflationary
stage the typical momentum of the relevant excitations is equal to the
Hubble parameter which for the potential (\ref{pot1}) is of the order
$H\simeq\sqrt{\lambda}M_P/\xi$.  This is much smaller than the cutoff
(\ref{EinsLambdainfl}) during inflation.  Similarly, the energy
density of the Universe during inflation, $\rho_I=\lambda M_P^4/\xi^2$,
is much smaller than $\Lambda^4$.

During reheating the scalar field $\phi$ oscillates with the amplitude
that decreases from $M_P/\sqrt{\xi}$ down to $M_P/\xi$.  Detailed
analysis in \cite{Bezrukov:2008ut} shows that typical momenta of
relativistic particles produced in this period are
$\sqrt{\lambda}\xi\bar\phi^2/M_P$ for the Higgs boson and $g\bar\phi$
for other light particles, where $g$ is the gauge coupling.  These are
again parametrically smaller than the corresponding cutoff value
(\ref{cutoffint}) with the suppression provided by the coupling
constants.\footnote{This is the only place where the presence of the
  gauge bosons make things more subtle.  As discussed in 
Appendix~\ref{app:bosons}, the cutoff may be comparable to the momentum of
  generated Higgs excitations during reheating.  This may affect the
  details of the description of reheating, but does not alter it
  qualitatively.} 

Finally, after the Universe reheats below the temperature
$T\sim{}M_P/\xi$, all relevant particle energies drop below the
small-field cutoff (\ref{cutoffsmall}).

We see that during the whole evolution of the Universe the relevant
energy scales are parametrically below the background dependent cutoff
$\Lambda(\bar\phi)$.  Clearly, this is a necessary requirement for the
validity of the semiclassical treatment of the model, which we thus
find fulfilled.  However, this requirement is not sufficient.  To
establish the sufficient conditions one has to analyse the loop
corrections to the tree-level picture.  We presently turn to this
task.

%%%%%%%%%%%%%%%%%%%%%%%%%%%%%%%%%%%%%%%%%%%%%%%%%%%%%%%%%%%%%%%%%%%%%%%%
\section{Size of quantum corrections, counterterms and all that}
\label{sec:quantum}

%%%%%%%%%%%%%%%%%%%%%%%%%%%%%%%%%%%%%%%%%%%%%%%%%%%%%%%%%%%%%%%%
\subsection{Effective field theory for inflation}
\label{sec:efftheory}

Of course, the theory (\ref{Lagr1}), or equivalently (\ref{Einstlagr})
is non-renormalisable.  This implies that the loop corrections will
generate infinite number of divergent counterterms.  Naively this
seems to imply the loss of predictive power.  We now argue that at
large fields which are relevant for inflation, $\chi>M_P$
($\phi>M_P/\sqrt\xi$), it is possible to consistently account for
quantum corrections in a way similar to that of effective field
theory.  The crucial property which enables to do this is the
approximate symmetry of the action in the inflationary region.  In the
Einstein frame it manifests itself as the symmetry under the shifts of
the scalar field, while in the Jordan frame it corresponds to the
scale invariance.

Let us analyse the structure of possible counterterms.  As usual, we
work in the background field formalism.  We stick to the Einstein
frame language where all non-trivial interactions are concentrated in
the scalar sector.  After canonically normalising the scalar field we
obtain the Lagrangian
\be
  \label{scallagr}
  \LL=\frac{(\d_\mu\chi)^2}{2}-U(\chi)
  \;,
\ee
where $U(\chi)$ has at large fields the generic form
\be
  \label{Ugenexp}
  U(\chi)=U_0\left(1+\sum_{n=1}^\infty u_n\e^{-n\chi/M}\right)
  \;.
\ee
For the concrete choice of the potential (\ref{pot1}) we have
\begin{equation}
  U_0=\frac{\lambda M^4_P}{4\xi^2}
  \;,\quad
  M=\frac{\sqrt{6}}{2} M_P
\end{equation}
and $u_n$ coincide with the Taylor coefficients of the function
$(1-x)^{2}$.  At $\chi\gg M$ the potential becomes constant giving
rise to the shift symmetry
\be
  \label{shift}
  \chi\mapsto\chi+\const
  \;.
\ee
If all coefficients $u_n$ were zero the shift symmetry would be exact
and the flatness of the potential would be preserved by radiative
corrections.  However, for viable inflation it is important that the
potential is not exactly flat due to the presence of exponential terms
and the shift symmetry is only approximate.  We will see nevertheless
that the fact that the symmetry is only weakly broken allows to
consistently take into account the quantum corrections during the
inflationary stage (cf.\ \cite{Cheung:2007st}).

Let us now split the field into smooth background $\bar\chi$ and
fluctuations $\delta\chi$ as in (\ref{split}).  The potential takes
the form,
\be
  \label{potsplit}
  U = U_0\left(1+\sum_{n=1}^\infty u_n\e^{-n\bar\chi/M}
    \sum_{k=0}\frac{1}{k!}\left[\frac{n\,\delta\chi}{M}\right]^k\right)
  = U_0\left(1+
    \sum_{k=0}^\infty\frac{1}{k!}\left[\frac{\delta\chi}{M}\right]^k
    \sum_{n=1}^\infty n^k u_n\e^{-n\bar\chi/M}\right)
  \;.
\ee
The Wilsonian effective action is given by integrating out the
perturbations $\delta\chi$.  Technically, this amounts to computing
all the loop diagrams generated by the interaction (\ref{potsplit})
without external legs of $\delta\chi$.  The background field
$\bar\chi$ in this procedure must be treated as classical.  Let us see
what kind of divergences this may produce.

Consider first the contributions to the effective potential. Clearly,
all divergences are proportional to the positive powers of the
exponent $\e^{-\bar\chi/M}$.  In other words, all counterterms can be
organised into a series of the form (\ref{Ugenexp}).  Thus they may be
absorbed by renormalisation of $U_0$ and the coefficients $u_n$.

Let us now discuss the loop divergences that require counterterms with
derivatives.  For example, at two and three loops one obtains
divergences of the following schematic form,
\setlength{\unitlength}{7ex}
\begin{align}
  \label{2loop}
  \frac{U_0u_n}{M^3}\e^{-n\bar\chi/ M}
  \raisebox{0.55ex}{\raisebox{-0.5\unitlength}{\begin{picture}(1.4,1)(-0.2,-0.5)
      \thicklines
      \put(0.5,0){\circle{1}}
      \put(0,0){\line(1,0){1}}
      \put(0,0){\circle*{0.2}} \put(1,0){\circle*{0.2}}
    \end{picture}}}
  \frac{U_0u_m}{M^3}\e^{-m\bar\chi/M}
  &\propto \frac{1}{\epsilon}\cdot \frac{U_0^2}{M^8}
  u_nu_m(\d_\mu \bar\chi)^2\e^{-(n+m)\bar\chi/M}
  \;,\\
  \label{3loop}
  \frac{U_0u_n}{M^4}\e^{-n\bar\chi/M}
  \raisebox{0.55ex}{\raisebox{-0.5\unitlength}{\begin{picture}(1.4,1)(-0.2,-0,5)
      \thicklines
      \put(0.5,0){\circle{1}}
      \put(0.5,0){\oval(1,0.5)}
      \put(0,0){\circle*{0.2}} \put(1,0){\circle*{0.2}}
    \end{picture}}}
  \frac{U_0u_m}{M^4}\e^{-m\bar\chi/M}
  &\propto \frac{1}{\epsilon}\cdot \frac{U_0^2}{M^8}
  u_nu_m\bigg(\frac{(\d^2 \bar\chi)^2}{ M^2}
  +\frac{(\d\bar\chi)^4}{M^4}\bigg)
  \e^{-(n+m)\bar\chi/M}
  \;,
\end{align}
where we assumed dimensional regularisation ($4-2\epsilon$ is the
space-time dimension) and omitted numerical
coefficients of order one and combinatorial factors. 
One makes two observations.  First, the
derivatives of $\bar\chi$ in these expressions are suppressed by
powers of the cutoff scale $M$ appearing from the differentiation of
the exponents.  Second, the $\bar\chi$-dependent coefficients in front
of the derivative terms are again proportional to positive powers of
the exponent $\e^{-\bar\chi/ M}$.  Thus we conclude that to absorb all
loop divergencies the Lagrangian (\ref{scallagr}) must be promoted to
\be
  \label{scalagr1}
  \LL=f^{(1)}(\chi)\frac{(\d_\mu\chi)^2}{2}-U(\chi)
  +f^{(2)}(\chi)\frac{(\d^2\chi)^2}{M^2}
  +f^{(3)}(\chi)\frac{(\d\chi)^4}{M^4}+\dotsb
  \;,
\ee 
where dots stand for terms with more derivatives.  Here the
coefficient functions are (formal) series in the exponent,
\be
  \label{genser}
  f^{(i)}(\chi)=\sum_{n=0}^\infty f_n^{(i)}\e^{-n\chi/M}
  \;,
\ee
where $f_n^{(i)}$ are numerical coupling constants.  It is
straightforward to convince oneself that this form of the Lagrangian
is preserved by quantum corrections, i.e.\ no new counterterms are
generated by any loops including those coming from the higher order
terms.\footnote{To avoid confusion we stress that this statement
  refers to the \emph{divergent} terms generated by loops.  Besides
  these there are of course finite quantum corrections which must be
  consistently taken into account.} In this sense we can speak about
renormalisation of the non-renormalisable Lagrangian (\ref{scalagr1})
with the coefficient functions (\ref{genser}).\footnote{Note that
  within purely scalar theory a somewhat more restricted choice with
  all $f^{(i)}(\chi)$, $i\geq 2$, vanishing at asymptotically large $\chi$ (this
  corresponds to taking $f_0^{(i)}=0$ for $i\geq 2$) is also stable
  under quantum corrections.  However, we do not impose this
  restriction as non-zero values of $f_0^{(i)}$ are inevitably
  generated once the theory is coupled to gravity.} Clearly, the
coefficient function $f^{(1)}(\chi)$ can be absorbed into redefinition
of the field leaving the kinetic term canonical without spoiling the
form (\ref{genser}) of the other terms.  Thus we will omit this
function in what follows.

So far we have been discussing explicitly purely scalar theory.
However, it is clear that the expansion formulated above holds when
the theory is coupled to gravity.  Indeed, this coupling respects the
asymptotic shift symmetry\footnote{Recall that we are working in the
  Einstein frame where coupling of $\chi$ to gravity is minimal.}
(\ref{shift}), and the violation of the latter still comes only in the
form of the exponents $\e^{-\chi/ M}$.  This guarantees that all
counterterms that appear in the perturbation theory\footnote{We do not
  discuss possible non-perturbative gravitational effects, which are
  model dependent and can
  be exponentially suppressed~\cite{Kallosh:1995hi}.} can be
arranged into shift-invariant operators multiplied by the coefficient
functions of the form (\ref{genser}).

The above statement remains true also upon inclusion of other fields
in the theory so far as their coupling to $\chi$ obeys the asymptotic
shift symmetry.  To illustrate this point let us consider a fermion
field with $\chi$-dependent mass term,
\be
  \label{chipsi}
  \LL_Y=m(\chi)\bar\psi\psi
  \;.
\ee
The asymptotic shift symmetry implies
\be
  \label{mshift}
  m(\chi)=\sum_{n=0}^{\infty}m_n\,\e^{-n\chi/ M}
\ee
at large $\chi$.  The divergent part of the one-loop contribution of
this fermion into the scalar potential is
\be
  \Delta U_\psi^{\mathrm{div}} \propto
  \frac{1}{\epsilon}\cdot [m(\chi)]^4
  \;.
\ee
Clearly, this has the structure (\ref{Ugenexp}) and is absorbed by the 
redefinition of $U_0$ and $u_n$.  Note that in the case when the
fermion remains massive at large $\chi$ (i.e. $m_0\neq 0$) 
the structure
(\ref{Ugenexp}) is also shared by the finite part of the loop,
\be
  \label{Upsifin}
  \Delta U_\psi^{\mathrm{fin}} \propto
  [m(\chi)]^4\,\ln[m(\chi)/\mu]
  \;,
\ee
where $\mu$ is the normalisation point, meaning that this correction
need not be considered separately and may be included in the general
expression (\ref{Ugenexp}) for the inflaton potential. 
The asymptotic shift symmetry of the inflaton
couplings to other fields naturally arises in the SM Higgs inflation
model \cite{Bezrukov:2008ej}.  It corresponds to the asymptotic
invariance of the original Jordan frame action (\ref{eq:TheAction})
under the scale transformations $\Phi(x)\to\lambda^D\Phi(\lambda x)$,
where $D$ is the canonical dimension of the field $\Phi$.

At first sight it seems that the presence of infinite number of
coupling constants $u_n$, $f^{(i)}_n$ implies the loss of predictive
power.  In fact, this is not the case.  At large values of $\chi$ the
exponent $\e^{-\chi/M}$ 
is small.  Thus requiring only finite accuracy, we can keep
only finite number of terms in the exponential series.  Also being
interested in characteristic momenta lower than the cutoff $ M$, one
can neglect the higher-derivative terms.\footnote{Or keep a finite
  number of them which gives rise to the expansion in $E/M$, where $E$
  is the energy scale of interest.  We are not going to discuss this
  expansion as it is completely standard.}  Then the theory is
determined by a finite number of parameters and (to this accuracy) the
predictive power is recovered. This is similar to the situation in
the standard EFT. From the practical point of
view the expansion in $\e^{-\chi/M}$ translates into the expansion in
the slow-roll parameters
\begin{align}
  \label{epsilon}
  \varepsilon &\equiv
  \frac{M^2_P}{2}\left(\frac{1}{U}\frac{dU}{d\chi}\right)^2\approx
  \frac{1}{3}u_1^2\e^{-2\chi/M}
  \;,\\
  \label{eta}
  \eta &\equiv\frac{M^2_P}{U}\frac{d^2U}{d\chi^2}\approx
  \frac{2}{3}u_1\e^{-\chi/M}
  \;.
\end{align}
This means that the inflationary predictions can be sensibly
calculated in the model, provided sufficient amount of the
coefficients $u_n$ is fixed form the observations (fixing just $u_1$
is enough for all modern applications).  For example, the well-known
property of inflation with the potential (\ref{Ugenexp}), which
follows from (\ref{epsilon}), (\ref{eta}), is the suppression of the
$\varepsilon$-parameter compared to $\eta$,
$\varepsilon\approx3\eta^2/4$.  Thus fixing $\eta\simeq-0.015$ from
the tilt of the spectrum of scalar perturbations yields the prediction 
for the tensor
to scalar perturbation ratio $r=16\varepsilon\simeq0.003$
\cite{Bezrukov:2007ep} (unfortunately, too small to be observed with
modern experimental techniques).  The results of this section show
that this prediction is robust under radiative corrections.

Let us stress once more that the crucial property which has enabled us
to develop the above EFT of inflation is the asymptotic shift symmetry
of the inflaton action.  This property naturally appears in the
Einstein frame and is preserved by quantum corrections with the
standard renormalisation prescriptions.  In this approach the original
Jordan frame action (\ref{eq:TheAction}) appears merely as a
convenient shorthand representation which ensures the asymptotic shift
symmetry of all inflaton couplings at tree level.  The suitable
language for the analysis of the quantum aspects of the theory is
provided by the Einstein frame.

It is worth comparing this approach to that of
Refs.~\cite{Barvinsky:2008ia,DeSimone:2008ei,
  Barvinsky:2009fy,Barvinsky:2009ii} and prescription II of
\cite{Bezrukov:2008ej,Bezrukov:2009db}.  There the loop corrections
are evaluated in the Jordan frame.  This calculation produces, among
others, the following contribution into the inflaton potential,
\be
  \label{DVJ}
  \Delta V^J\propto \phi^4\ln[\phi/\mu]
  \;,
\ee
where $\phi$ stands, as usual, for the Jordan frame field.  This
breaks the scale invariance at large $\phi$, and as a consequence
breaks the asymptotic shift symmetry in the Einstein frame.  The
reason is that choosing the standard renormalisation prescription with
fixed normalisation point $\mu_J$ in the Jordan frame corresponds to a
field-dependent normalisation point $\mu_E$ in the Einstein frame (see
discussion in \cite{Bezrukov:2008ej}),
\be
  \mu_E=\mu_J/\Omega(\chi)
  \;.
\ee
This relation is, of course, nothing but a manifestation of the
conformal anomaly.  Vice versa, fixing the Einstein frame
normalisation point leads to a field-dependent $\mu_J$ which
eliminates the logarithmic factor in (\ref{DVJ}) and preserves the
asymptotic scale invariance of the Jordan frame action.  This matches
with the fact that fixed $\mu_E$ preserves the asymptotic shift
symmetry in the Einstein frame.  Thus the choice of fixed $\mu_E$ is
favoured by the EFT framework developed in this section which
contains the asymptotic shift symmetry as the key ingredient.

%%%%%%%%%%%%%%%%%%%%%%%%%%%%%%%%%%%%%%%%%%%%%%%%%%%%%%%%%%%%%%%%
\subsection{Connection with the low energy physics}
\label{sec:lowhighconnection}

We have seen that the inflationary physics in the model at hand up to
a given order in the slow-roll parameters is determined by a finite number
of coupling constants.  It is important to understand if these
constants can be related to the observables measured at small energies
(i.e.\ at small values of the field) in the modern collider
experiments.  It turns out that the answer to this question cannot be
given within the EFT picture and depends on the
UV-completion.

One way to see the problem is to notice that establishing the desired
relation implies, in particular, the knowledge of the scalar potential
$U(\chi)$ in the whole range from $\chi=0$ to $\chi=\infty$.  Let us
assume that this potential is known at tree level and estimate quantum
corrections to it.  We concentrate on the divergent parts which
require introduction of counterterms in the bare action.  The
expansion of the bare potential in the background field method has the
form,
\be
  \lambda U(\bar\chi+\delta\chi)=\lambda\left(U(\bar\chi)+
    \frac{1}{2}U''(\bar\chi)(\delta\chi)^2
    +\frac{1}{3!}U'''(\bar\chi)(\delta\chi)^3+
    \frac{1}{4!}U^{(IV)}(\bar\chi)(\delta\chi)^4+\dotsb\right)
  \;,
\ee
where for convenience we have singled out explicitly the overall
coupling $\lambda$.  Computing loop corrections in, say, cutoff
regularisation scheme one generates the divergences of the form:
\begin{align}
  \label{onel}
  \text{in one loop:} &\quad
  \lambda U''(\bar\chi) \bar\Lambda^2,\ 
  \lambda^2(U''(\bar\chi))^2\log\bar\Lambda
  \;,\\
  \label{twol}
  \text{in two loops:} &\quad
  \lambda U^{(IV)}(\bar\chi) \bar\Lambda^4,\ 
  \lambda^2 (U''')^2 \bar \Lambda^2,\
  \lambda^3 U^{(IV)} (U'')^2(\log\bar\Lambda)^2
  \;,
\end{align}
where $\bar\Lambda$ is the loop cutoff.\footnote{It is worth stressing
  that $\bar\Lambda$ is just a technical parameter of the
  regularisation procedure and thus need not coincide with the
  tree-level unitarity violation scale found in Sec.~\ref{sec:scale}.
  In particular, one is free to choose $\bar\Lambda$ to depend or not
  on the background.} Similar results would be obtained in the
Pauli--Fierz regularisation.  According to the standard rules we have
to add corresponding counterterms to the bare Lagrangian in order to
absorb the divergences.  But the important point is that these
counterterms have a different functional dependence on the background
than the original potential.  So we cannot really absorb them in the
redefinition of the coupling constants.  Moreover, there is no natural
hierarchy between the lower and higher loop contributions.  In this
situation it is impossible to keep the radiative corrections under
control for all values of the fields.

The important role in this reasoning is played by power-law
divergences.  This type of divergences is particularly affected by the
UV-completion.  Thus one concludes that the form of the potential at
the intermediate values of $\chi$ is sensitive to the high-energy
physics.  To make this statement more precise let us consider a
UV-completion which involves a heavy particle with the mass $m(\chi)$
depending on the scalar field value.  At large values of the field the
$\chi$-dependence of the mass has to obey the asymptotic shift
symmetry (\ref{mshift}), but no further restrictions can be imposed on
it without the detailed knowledge of the UV theory.  Then, even if the
mass is made much higher than the scale of interesting processes and
at tree level the particle can be integrated out, on the one loop
level it will contribute the Coleman--Weinberg potential of the form
(\ref{Upsifin}).  In general this leads to large modification of the
potential, which is uncontrollable by the EFT.  Thus we conclude that
the inflationary and the present day physics cannot be connected
without specific assumptions about the UV completion.  (Still, both
can be separately described within the EFT framework.)

A possible choice of the additional assumption, which was effectively
adopted in \cite{Bezrukov:2008ej,Bezrukov:2009db}, is that the UV
theory is such that the power-law divergences must be discarded
altogether.\footnote{The physics behind this hypothesis is discussed in
\cite{Shaposhnikov:2008xb,Shaposhnikov:2008xi} and is associated with
exact, but spontaneously broken quantum scale invariance.}  It
implies that the UV theory does not contain any heavy elementary
particles beyond the SM (or alternatively, the effect of heavy degrees
of freedom on the low-energy physics completely cancels out).
Technically, this assumption is implemented by the use of the
scale-free minimal subtraction scheme based on dimensional
regularisation in all loop computations.  Then we see from
(\ref{onel}), (\ref{twol}) that the remaining (logarithmic)
divergences are suppressed by an extra power of the coupling constant
$\lambda$ for each additional loop.  So one can arrange the
perturbative expansion in such a way that the divergences are absorbed
order by order in $\lambda$.  Namely one writes the potential as a
formal series in $\lambda$,
\be
  U(\chi)=\lambda U_1(\chi)+\lambda^2 U_2(\chi)
  +\lambda^3 U_3(\chi)+\dotsb
  \;.
\ee
Then the divergences originating from $U_i$ feed in only into terms of
order $i+1$ and larger in $\lambda$.  The functions $U_j$, $j\geq i+1$
can be chosen in such a way as to absorb these divergences.  

Note that there are two physically distinct types of ambiguities
associated with this procedure. First, the functions $U_i$, $i\geq 2$,
are determined up to the finite part of the counterterms generated by
the loops with $U_j$, $j<i$, insertions. This uncertainty is
unavoidable in any quantum treatment of the model. The second source
of uncertainty is the freedom in the choice of the
tree-level action. Indeed, all functions $U_i$ can, in principle,
be modified by an addition of an arbitrary function. This ambiguity
eventually amounts
to defining the model and
must be fixed by some physical considerations.

Let us illustrate this point by considering the function $U_2$. 
Representing it in the form
\be
  \label{U2}
  U_2(\chi)=c_{21}\cdot \big(U_1''(\chi)\big)^2+\tilde U_2(\chi)
  \;,
\ee
we see that the one-loop divergence coming from $U_1$ is absorbed into
the renormalisation of the coefficient $c_{21}$. This coefficient is
thus promoted to an independent coupling constant that must be fixed
from measurements. Its a priori unknown value represents an
uncertainty of the first type. On the other hand, the
unrenormalised contribution $\tilde U_2(\chi)$ is related to the second type
of uncertainties. Its choice makes part of the
definition of the model. For example, 
it can be consistently put to
zero.

It appears that the same rearrangement can be done with the
divergences appearing in the kinetic term and higher derivative terms.
When $\chi$ is coupled to other fields, say, fermions and gauge
fields, one has to promote all functions of $\chi$ in the action to a
formal series in all coupling constants including the Yukawa and gauge
couplings.  Then, at least formally, the physics to any finite order
in the coupling constants is determined by finite number of
parameters.  Note though that to make predictions in this approach one
still needs an additional principle to fix the ambiguities
in the coefficient functions of the formal
series, such as, e.g., the choice of $\tilde U_2(\chi)$ in (\ref{U2}).

In \cite{Bezrukov:2008ej,Bezrukov:2009db} this ambiguity was fixed by
choosing the finite parts of all the coefficient functions
beyond $U_1$ to be zero (i.e. the $\overline{MS}$ subtraction scheme was
used).  Let us estimate what kind of uncertainty is introduced in the
bounds on the Higgs mass obtained in
\cite{Bezrukov:2008ej,Bezrukov:2009db} by variations of the
subtraction scheme.\footnote{In the terminology of 
Refs.~\cite{Bezrukov:2008ej,Bezrukov:2009db} we are considering the
renormalisation prescription I.} 
This uncertainty is represented by the finite part
of the counterterms which we thus need to estimate. For this
end we consider the bare potential
\be
U_B= \lambda U_1+\left(\frac{1}{\epsilon} + c\right) 
\frac{\lambda^2 U_1''^2}{64\pi^2}~,
\ee
where we wrote explicitly the loop suppression factor and $c$ is an
arbitrary constant. The variation of $c$ does not change the predictions
for inflation, since the contribution of the finite part of the
counter-term is negligible in the inflationary region. However, this
term modifies the relation between the low energy Higgs mass $m_H$ and the
high energy scalar self-coupling, leading to the change in the lower and
upper Higgs mass bounds by an amount
\be
  \delta m_H \simeq \frac{9 c}{64\pi^2}\frac{m_H^3}{v^2}
  \;,
\ee
where $v=246.22$~GeV is the SM Higgs field vacuum expectation
value.
Numerically, for $c \sim 1$, this change is $\sim 0.5$ GeV for the lower
bound on the Higgs mass ($\simeq 126$ GeV) and is about $2$ GeV for the
upper bound ($\simeq 194$ GeV). This is smaller than other
uncertainties, related, e.g.\ to the experimental error in the mass of
the top quark \cite{Bezrukov:2009db}. However, if $c$ is large
(say, $|c| >10$), the uncertainties introduced by the finite part of
counter-terms will dominate the error bars.

%%%%%%%%%%%%%%%%%%%%%%%%%%%%%%%%%%%%%%%%%%%%%%%%%%%%%%%%%%%%%%%%%%%%%%%%
\section{Generalisation}
\label{sec:general}

Here we discuss the application of the EFT ideas developed in
Sec.~\ref{sec:efftheory} to a general class of inflationary models
having the property of asymptotic shift symmetry.  Let us start from a
scalar theory with the potential
\be
  \label{UF}
  U(\chi)=U_0\big(1+u_1 F(\chi/M)+\dotsb\big)
  \;,
\ee
where $F(x)$ is a given function with the property
\be
  \label{asymp}
  F(x)\to 0  \quad\text{at}\quad  x\to +\infty
\ee
together with its derivatives, and $u_1$ is a coefficient of order
one.  Potentials of this class appear in many inflationary models, see
e.g.~\cite{Lyth:1998xn} and references therein.  Inflation happens at
$\chi > M$.  Dots in (\ref{UF}) stand for the terms which are yet to
be determined and which, as we will see, are required for consistency.
The case considered in Sec.~\ref{sec:efftheory} corresponds to
$F(x)=\e^{-x}$.  Another example to have in mind is
\be
  \label{powerlaw}
  F(x)=\frac{1}{x^\alpha}
  \;,\quad \alpha > 0
  \;.
\ee 
The expansion of the potential in the background field formalism is
\be
  \label{Uexp}
  U(\bar\chi+\delta\chi)=U_0\bigg(1+u_1\sum_{k=0}^\infty
  \frac{F^{(k)}(\bar\chi/M)}{k!}\left(\frac{\delta\chi}{M}\right)^k
  +\dotsb\bigg)
  \;,
\ee
where $F^{(k)}$ is the $k$-th derivative of the function $F(x)$.
One observes that the loop integrals calculated using
(\ref{Uexp}) generate divergences proportional to the products of
derivatives of~$F$,
\be
  F^{(k_1)}\bigg(\frac{\bar\chi}{M}\bigg)
  F^{(k_2)}\bigg(\frac{\bar\chi}{M}\bigg)
  \dotsm
  F^{(k_n)}\bigg(\frac{\bar\chi}{M}\bigg)
  \;.
\ee
In general, all possible combinations of this type appear with the
only restriction that the sum
$$
  k_1+k_2+\dotsb+k_n
$$
is even, following from the fact that all the lines of the
perturbation $\delta\chi$ must be closed into loops.  This requires
inclusion of the corresponding counterterms in the bare action.  Thus
we conclude that (\ref{UF}) must be extended to the following formal
series,
\be
  \label{UFF}
  U(\chi)=U_0\bigg(1+\sum_{\begin{smallmatrix}
      k_1,k_2,\dotsc,k_n \\
      k_1+k_2+\dotsb+k_n - \mathrm{even}\end{smallmatrix}}
  u_{k_1,k_2,\dotsc,k_n}
  F^{(k_1)}\bigg(\frac{\chi}{M}\bigg)
  F^{(k_2)}\bigg(\frac{\chi}{M}\bigg)
  \dotsm
  F^{(k_n)}\bigg(\frac{\chi}{M}\bigg)\bigg)
  \;,
\ee
where $u_{k_1,k_2,\dotsc,k_n}$ are arbitrary coefficients (coupling
constants).  Similar extension must be performed with all coefficient
functions appearing in the Lagrangian, i.e.\ those multiplying the
terms with
%higher
derivatives of $\chi$ and the interactions of $\chi$ with other
fields.  It is straightforward to convince oneself that the resulting
form of the Lagrangian is stable under (perturbative) quantum
corrections.  Note that for particular choices of the function $F(x)$
some subsets of the terms in the series (\ref{UFF}) may collapse to a
single term.  For example, this happens for the choice $F(x)=\e^{-x}$
when (\ref{UFF}) becomes a simple series in the powers of the exponent
considered in Sec.~\ref{sec:efftheory}.

A comment is on order.  It appears natural to identify the scale $M$
with the cutoff in the loop integrals.  In this case the size of the
loop corrections to the coefficients $u_{k_1,k_2,\dotsc,k_n}$ is of
the order $O[(U_0/M^4)^{n-1}]$, which can be obtained by iterative use
of (\ref{Uexp}) in the perturbative diagrams.  This gives the bound on
the natural value of these coefficients:
\be
  \label{hier}
  u_{k_1,k_2,\dotsc,k_n} \gtrsim
  \left(\frac{U_0}{M^4}\right)^{n-1}
  \;.
\ee
Note that the ratio on the r.h.s.\ of this bound is smaller that one.
This follows from the requirement that the energy density during
inflation must be smaller than the cutoff to the fourth power.  Thus
the choice of the coefficients $u_{k_1,k_2,\dotsc,k_n}$ saturating the
bound (\ref{hier}) corresponds to hierarchically small values of the
coefficients with higher $n$.

One may wonder if the expression (\ref{UFF}) with its infinite number
of terms is of any use.  The answer is yes, provided the subsequent
terms in the series vanish at $\chi\to +\infty$ faster than the
previous ones.  Then at the inflationary epoch corresponding to large
$\chi$ one can account for them in the framework of an EFT expansion
analogous to that developed in Sec.~\ref{sec:efftheory}.  Loosely
speaking this requires that the function $F(x)$ is such that
\be
  F^{(k+1)}(x)< F^{(k)}(x)
  \quad\text{at}\quad x\to +\infty
  \;.
\ee
This restriction is rather mild and is satisfied by reasonable
choices\footnote{In this sense the choice $F(x)=\e^{-x}$ is a limiting
  case, $F^{(k+1)}= F^{(k)}$.  Additional property that enables to
  develop the EFT description in this case is the simplification of
  the series (\ref{UFF}) pointed above.}  of $F(x)$.  Additionally,
the higher terms in the series (\ref{UFF}) may be suppressed if the
coefficients $u_{k_1,k_2,\dotsc,k_n}$ are chosen to obey the hierarchy
corresponding to the saturation of the bound (\ref{hier}).

Let us see explicitly how the EFT picture works for the power-law
choice (\ref{powerlaw}).  Substituting the derivatives into
(\ref{UFF}) and combining the terms of the same form together we
arrive at the following double series,
\be
  \label{Ups}
  U(\chi)=U_0\left(
    1+\sum_{n=1,\,m=0}^\infty u_{n,m}
    \left(\frac{M}{\chi}\right)^{n\alpha+2m}
  \right)
  \;.
\ee
Keeping only the leading term in the series one obtains certain
expressions for the inflationary observables \cite{Lyth:1998xn}.  In
particular, for the slow-roll parameters one has
\begin{align}
  \varepsilon &= \frac{\alpha^2M_P^2}{2M^2}\, u_{1,0}^2\,
  \bigg(\frac{M}{\chi}\bigg)^{2\alpha+2}
  \;,\\
  \label{etapow}
  \eta &= \frac{\alpha (\alpha+1)M_P^2}{M^2}\,u_{1,0}\,
  \bigg(\frac{M}{\chi}\bigg)^{\alpha+2}
  \;.
\end{align}
To be concrete, we assume that $M$ is of order $M_P$ and
$u_{1,0}=O(1)$.  Then we obtain the relation
\be
  \label{epseta}
  \varepsilon\simeq \eta^{\frac{2\alpha+2}{\alpha+2}}
  \;.
\ee 
Let us use (\ref{Ups}) to estimate the size of quantum corrections to
the inflationary observables.  The first subleading terms in the
series (\ref{Ups}) are
\be
  \label{sublead}
  u_{1,1}\left(\frac{M}{\chi}\right)^{\alpha+2}
  \quad\mathrm{and}\quad
  u_{2,0}\left(\frac{M}{\chi}\right)^{2\alpha}
  \;.
\ee
According to the above discussion the natural size of the coefficients
$u_{1,1}$, $u_{2,0}$ generated by loop effects is
\be
  u_{1,1}\sim 1 \;,\quad u_{2,0}\sim \frac{U_0}{M^4}
  \;.
\ee
We now show that for a realistic inflationary model the second term in
(\ref{sublead}) is negligible compared to the first.  From the COBE
normalisation we have $U_0/M^4\ll \varepsilon$ and thus
\be
  \label{2nd}
  u_{2,0}\left(\frac{M}{\chi}\right)^{2\alpha}\ll
  \varepsilon\,\eta^{\frac{2\alpha}{\alpha+2}}
  \simeq\eta^\frac{4\alpha+2}{\alpha+2}
  \;,
\ee 
where we have used the relations (\ref{etapow}), (\ref{epseta}).  On
the other hand,
\be
  u_{1,1}\left(\frac{M}{\chi}\right)^{\alpha+2}\sim \eta
  \;,
\ee
which is indeed larger than (\ref{2nd}) for any positive $\alpha$ and
$\eta\ll 1$.  Thus the main correction comes from the first term in
(\ref{sublead}).  Its relative size compared to the leading term is
$(M/\chi)^2$.  Expressed via the slow-roll parameter this translates
into $\eta^{2/(\alpha+2)}$.  Note that the correction behaves as a
fractional power of $\eta$.

It is worth mentioning that the modification of the Higgs inflation
proposed recently in \cite{Lerner:2010mq} belongs to the class
of models considered in this section.  It corresponds to the power-law
choice (\ref{powerlaw}) for the function $F(x)$ and $\alpha=2$.  Note
though that in the case of \cite{Lerner:2010mq} the scale $M$ is much
lower than\footnote{Namely, $M=M_P/\sqrt{\xi}$, where $\xi\gg 1$ is
  the coefficient of the non-minimal coupling of the inflaton to
  curvature in the Jordan frame.} $M_P$, so the formulas of the
previous paragraph should be appropriately modified when applied to
this case.

%%%%%%%%%%%%%%%%%%%%%%%%%%%%%%%%%%%%%%%%%%%%%%%%%%%%%%%%%%%%%%%%%%%%%%%%
\section{Conclusions}
\label{sec:conclusions}

In this paper we addressed the sensitivity of the Higgs inflation
scenario proposed in \cite{Bezrukov:2007ep} to the details of the UV
completion of the theory.  We determined the cutoff of the
theory which we identified with the scale of violation of the
tree-level unitarity.  With this definition the cutoff depends on the
background value of the inflaton (Higgs) field and we found that it is
larger than the characteristic energy scale involved in the physical
processes during the whole history of the Universe. 

For clarity we
concentrated on a simplified toy model obtained from the model of
\cite{Bezrukov:2007ep} by suppressing fermion and gauge fields. This
captures the main features, namely the background dependence of the
cutoff and the properties of the quantum corrections to the potential.
As discussed in Appendix \ref{app:fermions}, 
inclusion of the SM fermions does not affect our results. 
The situation is slightly more complicated when the gauge bosons are
taken into account. The cutoff in the gauge sector is
lower than in the pure inflaton theory, see Appendix
\ref{app:bosons}. Still, it is parametrically higher than the relevant
energy scales at inflation and during the subsequent evolution
(modulo subtleties at the beginning of the reheating epoch, see
Appendix \ref{app:bosons}). 

We analysed the quantum corrections to the tree-level Lagrangian of
the theory and formulated the assumptions about the UV completion that
allow to keep these corrections under control.  At the proper
inflationary stage the sufficient conditions are concisely formulated
as the requirement of asymptotic symmetry of the theory at large
values of the inflaton field.  Depending on whether one works in
Jordan or Einstein frame, the corresponding symmetry is either scale
invariance or the symmetry under the shifts of the field.  This
asymptotic symmetry allows to arrange the quantum corrections to the
inflaton potential in an infinite series where subsequent terms are
suppressed by a small parameter which is physically identified with (a
power of) the slow-roll parameter $\eta$.  Besides, the Lagrangian of
the low-energy theory contains standard EFT contributions with higher
derivatives of the inflaton field considered in
\cite{Cheung:2007st,Weinberg:2008hq}.  Thus our results extend the EFT
approach to inflation to the case of the inflaton potential.

The approach developed in this paper is general and we showed how it
can be applied to a wide class of inflationary models with asymptotic
shift (or scale) symmetry.  In particular, the modification of the
Higgs inflation model proposed in \cite{Lerner:2010mq} belongs to this
class.  On the other hand, our method does not apply to the ``new
Higgs inflation'' of \cite{Germani:2010gm,Germani:2010ux} which does
not seem to possess any asymptotic shift or scale invariance, and thus
requires a separate study of stability against radiative
corrections.

We also analysed in the context of the model of \cite{Bezrukov:2007ep}
under which conditions the parameters of the inflationary EFT can be
related to the observables measured at low energies.  We showed that
this requires further assumptions about the UV completion.  We
considered the assumption that the UV completion is such that all
power-law divergences appearing in the loop diagrams must be
discarded.  Physically, this amounts to the statement that the effects
of possible heavy degrees of freedom present in the full theory on the
low-energy physics completely cancel out.  We demonstrated that this
(admittedly, strong) assumption is sufficient to establish the link
between inflation and the low-energy physics once the tree-level
action of the model is fixed.

Let us conclude with the following remark.  The approach we adopted in
this paper is that of the effective field theory.  Thus the asymptotic
shift symmetry that was crucial for us to develop the consistent
perturbation scheme for inflationary calculations was just assumed: we
did not address the question how it appears at
the level of UV-complete theory. One may speculate in this connection
that the truly fundamental property is the scale invariance of the
Jordan frame action at large values of the inflaton (which is, of
course, equivalent to the Einstein frame shift symmetry). 
This may indicate that the UV-complete theory has an exact, but
spontaneously broken quantum scale invariance, relevant also for gauge
hierarchy, cosmological constant, and dark energy problems
\cite{Shaposhnikov:2008xb,Shaposhnikov:2008xi}. This may also be
related to the existence of a scale invariant UV fixed point. Recently
similar ideas were expressed in \cite{Ferrara:2010in}. 
It would be interesting to understand the connection between this
work and the results of the present paper.

%%%%%%%%%%%%%%%%%%%%%%%%%%%%%%%%%%%%%%%%%%%%%%%%%%%%%%%%%%%%%%
\paragraph*{Acknowledgements}
%%%%%%%%%%%%%%%%%%%%%%%%%%%%%%%%%%%%%%%%%%%%%%%%%%%%%%%%%%%%%%

We thank Andrei Barvinsky, Dmitry Gorbunov, Oriol Pujol\`as, Riccardo
Rattazzi, Alexei Starobinsky and Andrea Wulzer for helpful
discussions.  This work was supported in part by the Swiss National
Science Foundation (A.M. and M.S.), Tomalla Foundation (S.S.), RFBR grants
08-02-00768-a and 09-01-12179 (S.S.)  and the Grant of the President
of Russian Federation NS-1616.2008.2~(S.S.).

%%%%%%%%%%%%%%%%%%%%%%%%%%%%%%%%%%%%%%%%%%%%%%%%%%%%%%%%%%%%%%%%%%%%%%%%
\appendix

%%%%%%%%%%%%%%%%%%%%%%%%%%%%%%%%%%%%%%%%%%%%%%%%%%%%%%%%%%%%%%%%%%%%%%%%
\section{Fermions}
\label{app:fermions}

Let us analyse the effect of the fermion field with the Lagrangian in
the Jordan frame,
\be
  \label{Yuk1}
  \LL_Y^J=i\bar\psi\slashed\partial\psi+y\phi\bar\psi\psi
  \;,
\ee
where $y$ is the coupling constant.  After conformal transformation
(\ref{conftrans}) with the appropriate rescaling of the fermion
\begin{equation}
  \psi\mapsto\Omega^{3/2} \psi
  \;,
\end{equation}
the interaction in the Einstein frame becomes
\be
  \label{Yuk2}
  \LL^E_Y=y\frac{\phi(\chi)}{\sqrt{1+\xi\phi^2(\chi)/M_P^2}}\,\bar\psi\psi
  \;.
\ee
With the help of the relation (\ref{varphiphi}) it is easy to
reproduce the results of the Sec.~\ref{sec:scale} in all three regions
of the background field.

\begin{itemize}
\item $\bar\phi,\bar\chi\ll M_P/\xi$.  Using (\ref{eq:philowchiexpansion}) we
  get
  \begin{equation}
    \Lambda_{(n)} \sim \frac{M_P}{\xi} y^{-1/(n-4)}
    \;,
  \end{equation}
  which coincides (apart from the obvious change of the coupling constant)
  with (\ref{eq:8}).
\item Intermediate region $M_P/\xi\ll\bar\phi\ll M_P/\sqrt{\xi}$.  In this
  region the relation between the $\phi$ and $\chi$ fields
  is
  \be
    \label{phichi1}
    \phi \approx
    \left(\frac{2}{3}\right)^{1/4}\sqrt{\frac{M_P\chi}{\xi}}
    \;,
  \ee
  and thus (\ref{Yuk2}) takes the form
  \be
    \LL_Y^E\approx y\left(\frac{2}{3}\right)^{1/4}\sqrt{\frac{M_P\chi}{\xi}}\,
    \bar\psi\psi
    \;.
  \ee
  Clearly, expansion of this expression in the perturbations
  $\delta\chi$ produces an infinite series of higher-order
  operators of the form
  \[
    y\left(\frac{2}{3}\right)^{1/4}\sqrt{\frac{M_P}{\xi\bar\chi}}\cdot
    \frac{(\delta\chi)^n\bar\psi\psi}{\bar\chi^{n-1}}
    \;.
  \]
  This shows that for moderately small Yukawa couplings the scale
  suppressing higher interactions is essentially equal to $\bar\chi$
  which gives the cutoff
  \be
    \Lambda\simeq\bar\chi
    \;.
  \ee
  Given the relation (\ref{phichi1}) this coincides with the
  expression (\ref{cutoffint}) obtained in the Jordan frame.
\item $\bar\phi\gg M_P/\sqrt{\xi}$.  Using the relation (\ref{chiphi2})
  we obtain the exponential interaction
  \begin{equation}
    \frac{yM_P}{\sqrt\xi}\bigg(1-\frac{1}{2}\e^{-\frac{2\chi}{\sqrt{6}M_P}}+\ldots
\bigg)
\bar\psi\psi
    \;,
  \end{equation}
  which upon expanding in the excitations $\delta\chi$ yields the Planck
  mass cutoff (\ref{EinsLambdainfl}).
\end{itemize}
Thus we conclude that the 
addition of fermions does not change the value of the cutoff.

%%%%%%%%%%%%%%%%%%%%%%%%%%%%%%%%%%%%%%%%%%%%%%%%%%%%%%%%%%%%%%%%%%%%%%%%
\section{Gauge bosons}
\label{app:bosons}

The addition of the gauge bosons is also most simple to analyse in the
Einstein frame.  As the vector fields do not change under the
conformal transformation (\ref{conftrans}), the only change is in the
gauge-Higgs interactions.  For example, in the unitary gauge the only
change is in the mass terms of the gauge bosons,
\begin{equation}
\label{eq:B1}
  g^2 \phi^2 A_\mu A_\mu \rightarrow
  g^2 \frac{\phi(\chi)^2}{\Omega^2(\chi)} A_\mu A_\mu
  \;,
\end{equation}
where $A_\mu$ is a gauge boson, and $g$ is the weak coupling constant.
Full action in arbitrary gauge can be obtained similarly to the chiral
electroweak model and is described in detail in~\cite{Bezrukov:2009db}.

We see that at nonzero background the Higgs field excitations
$\delta\chi$ interact with the gauge bosons weaker than in 
the normal Higgs
model. Thus the Higgs mechanism fails at the
inflationary and reheating epochs---the diagrams with the
perturbations of the Higgs field do not cancel the growth of the
amplitudes with non-abelian vector bosons.  Thus the gauge fields will
enter into strong coupling at the energy $m_A(\bar\chi)/g$, where
$m_A(\bar\chi)$ is the mass of the vector bosons.  Reading out the
expressions for $m_A(\bar\chi)$ from (\ref{eq:B1})
\cite{Bezrukov:2007ep,Bezrukov:2008ut,Bezrukov:2009db} we obtain the
(Einstein frame) cutoff in the gauge sector,
\be
  \Lambda_A(\bar\chi)\simeq\begin{cases}
    \bar\phi\simeq \sqrt\frac{M_P\bar\chi}{\xi}&
    \mathrm{at}~~\frac{M_P}{\xi}\lesssim\bar\chi\lesssim M_P\;,\\
    \frac{M_P}{\sqrt\xi} &\mathrm{at}~~ M_P\lesssim\bar\chi\;.
  \end{cases}
\ee
This is lower than the values (\ref{eq:cutoffintE}),
(\ref{EinsLambdainfl}) obtained in the pure inflaton model.  Still,
during inflation $\Lambda_A$ is parametrically higher than all the
characteristic energy scales discussed at the end of
Sec.~\ref{sec:scale}.  Also during reheating the momenta of the vector
bosons, which are of order $m_A$ \cite{Bezrukov:2008ut}, are safely
below $\Lambda_A$.  Some tension arises for the momenta of the Higgs
excitations, $p_{\mathrm{Higgs}}\sim \sqrt\lambda\xi\bar\phi^2/M_P$
\cite{Bezrukov:2008ut}, which may be larger than $\Lambda_A$ at the
beginning of the reheating period corresponding to
$\bar\phi\simeq{}M_P/\sqrt\xi$.  However, significant energy is
transfered to the relativistic Higgs excitations only at later stages.
Thus one does not expect this lower cutoff to alter qualitatively the
analysis of reheating in
\cite{Bezrukov:2008ut,GarciaBellido:2008ab}.

%%%%%%%%%%%%%%%%%%%%%%%%%%%%%%%%%%%%%%%%%%%%%%%%%%%%%%%%%%%%%%%%%%%%%%%%

\providecommand{\href}[2]{#2}\begingroup\raggedright\endgroup
%\bibliographystyle{JCAP-hyper}
%\bibliography{all,misc}

\end{document}